\documentclass[aps
floats,prd,nofootinbib,superscriptaddress,10pt]{revtex4-1}
\usepackage[utf8]{inputenc}
\usepackage{bm}
\usepackage{amsmath}
\usepackage{comment}
\usepackage{color}
\def\blue{\textcolor{black}}

\def\red{\textcolor{black}}
\def\cyan{\textcolor{black}}
\def\magenta{\textcolor{black}}
\usepackage{graphicx,amsmath,amsfonts,amssymb
}

\begin{document}

\title{Micro black holes formed in the early Universe and their cosmological implications
}

\author{Tomohiro Nakama}

\affiliation{Department of Physics and Astronomy, Johns Hopkins
     University, 3400 N.\ Charles St., Baltimore, MD 21218, USA}
     
     \affiliation{Institute for Advanced Study, The Hong Kong University of Science and Technology, Clear Water Bay, Kowloon, Hong Kong}
     
     \author{Jun'ichi Yokoyama}
     
     \affiliation{Research Center for the Early Universe (RESCEU),
Graduate School of Science, The University of Tokyo, Tokyo 113-0033, Japan}

\affiliation{Department of Physics, Graduate School of Science,
The University of Tokyo, Tokyo 113-0033, Japan and}

\affiliation{Kavli Institute for the Physics and Mathematics
of the universe (Kavli IPMU), WPI, UTIAS,
The University of Tokyo, 5-1-5 Kashiwanoha, Kashiwa 277-8583, Japan}

\begin{abstract}
High energy collisions of particles may have created tiny black holes in the early Universe, which might leave stable remnants instead of fully evaporating as a result of Hawking radiation. If the reheating temperature was sufficiently close to the fundamental gravity scale, which can be different from the usual Planck scale depending of the presence and properties of spatial extra-dimensions, the formation rate could have been sufficiently high and hence such remnants could account for the entire cold dark matter of the Universe.
\end{abstract}

\maketitle
\section{Introduction}
It is now widely accepted that black holes are formed as a result of gravitational collapse of stars. The masses of resultant black holes are \blue{larger than the} solar mass. In the early Universe, black holes of significantly smaller masses could have been formed through a variety of mechanisms (\blue{see Ref. \cite{Carr:2009jm} and references therein}), some of which will be mentioned in this paper. Interestingly, the mass of a black hole decreases due to Hawking radiation \cite{Hawking:1974sw}, which is particularly important for such small black holes.

The Hawking radiation is derived by treating matter fields quantum mechanically, while treating the space-time metric classically. When the mass of an evaporating black hole becomes comparable to the Planck scale, such a treatment would breakdown, and quantum gravitational effects would become relevant. Hence, the \blue{final} state of Hawking evaporation is unknown, and stable Planck-mass relics may be left over (see \cite{Chen:2014jwq} for a review). \blue{Torres et al. \cite{Torres:2013kda} argued} that imposing energy conservation alone could be sufficient to prevent complete evaporation. Such remnants could even be white holes \cite{Bianchi:2018mml,Rovelli:2018hbk}. 
A sufficient amount of these remnants can account for the cold dark matter \cite{MacGibbon:1987my}. A pre-inflationary phase dominated by black hole remnants is discussed in Ref. \cite{Scardigli:2010gm}. Black hole remnants which possibly arose before inflation \red{(see \cite{Sato:2015dga} for a review)} 
were likely to have been substantially diluted and hence they would not account for the entire dark matter \cite{Chen:2002tu}. 

The properties and hence formation of black holes can be significantly altered if spatial extra dimensions are present, which were introduced to solve the hierarchy problem in Ref. \cite{ArkaniHamed:1998rs}. Astrophysical as well as cosmological limits on such a framework are subsequently discussed in Ref. \cite{ArkaniHamed:1998nn}. The properties of black holes and limits on primordial black holes in this context are discussed in Ref. \cite{Argyres:1998qn}. See also Ref. \cite{Majumdar:2005ba} for black hole geometries and evolution of primordial black holes in a brane world cosmology. The phenomenology of large, warped, and universal extra dimensions is reviewed in \cite{Kribs:2006mq}. 

In such a framework, the fundamental gravity scale can be significantly smaller than the usual Planck energy. \magenta{Then} black holes could be formed at collider experiments, and this topic has been extensively discussed in the literature. See Refs. \cite{Gingrich:2006hm,Kanti:2008eq,Park:2012fe} for reviews of black holes at the Large Hadron Collider. Hypothetical stable micro black hole production
at a future 100 TeV collider is discussed in Ref. \cite{Sokolov:2016lba}.

Black \blue{hole} formation due to high-energy particle collisions is expected to have been efficient in the early Universe \cite{Conley:2006jg,Saini:2017tsz}. We explore \blue{the} possibility that such micro black holes survive today, as opposed to fully evaporating, to account for the entire cold dark matter. We also consider cases with extra dimensions. See also Refs. \cite{Barrau:2005zb,Suranyi:2010yt} for relevant discussions.

\section{Micro Black holes formed in the early Universe}

As mentioned above, black hole formation from high-energy particle collisions has been extensively discussed in the literature. 
See \magenta{also} Ref. \cite{Choptuik:2009ww} for a numerical simulation, and Ref. \cite{Yoshino:2005hi} for an analysis based on a superposition of two boosted Schwarzschild metrics. The formation depends on the properties of colliding particles such as charge \cite{Yoshino:2006dp} or spin \cite{Yoshino:2007ph}. First let us estimate the production rate assuming no extra dimensions. In the following, for a simple and crude estimation, we assume that a black hole is formed if two particles both with kinetic energy larger than the Planck energy collide with an impact parameter less than the Planck length $l_P$.


The number density of particles with kinetic energy above the Planck energy $E_P$ when the Universe is in thermal equilibrium at temperature $T$ is 
\begin{equation}
n\sim T^3\int_{x_P}^\infty x^2e^{-x}dx\sim T^3 x_P^2e^{-x_P}\sim l_P^{-3}\frac{T}{T_P}\exp\left(-\frac{E_P}{kT}\right),
\end{equation}
where $x=E/kT$, $x_P=E_P/kT$ and $T_P\blue{=E_P/k}$ is the Planck temperature, \magenta{with $k$ being the Boltzmann constant}. 
The probability $\Gamma$ of a particle colliding with another particle with kinetic energy above $E_P$ per unit time is $\Gamma=n\sigma v\sim t_P^{-1}(T/T_P)\exp(-E_P/kT)$, where we have assumed $\sigma\sim l_P^2$ for the cross section, $v\sim c$ for the relative velocity and $t_P$ is the Planck time. The energy density $\rho$ of radiation is $\rho= \blue{(\pi^2g_*/30)}E_Pl_P^{-3}(kT/E_P)^4$, and the Hubble parameter is $H\blue{=(4\pi^3g_*/45)^{1/2}} t_P^{-1}(kT/E_P)^2$. \blue{Here, $g_*$ denotes the effective number of relativistic degrees.} The energy density of relics which arise during the Hubble time at the reheating is $\rho_{\mathrm{rel}}\sim E_P n H^{-1} \Gamma$, and let us introduce $\beta\equiv \rho_{\mathrm{rel}}/\rho$. Neglecting the decrease in relativistic degrees of freedom for simplicity \cite{Nakama:2016enz}, which would not affect the following conclusions much, $\beta$ roughly grows in proportion to $T^{-1}$ by the matter-radiation equality: 
\begin{equation}
\beta_{\mathrm{eq}}\sim \frac{T}{T_{\mathrm{eq}}}\beta\sim \blue{\frac{30}{\pi^2g_*}\left(\frac{45}{4\pi^3g_*}\right)^{1/2}}\frac{kT/E_P}{kT_{\mathrm{eq}}/E_P}\left(\frac{kT}{E_P}\right)^{-4}\exp\left(-\frac{2E_P}{kT}\right).\label{betaeq}
\end{equation}
If the maximum temperature of the radiation-dominated Universe reached $kT\sim 0.01 E_P$, then $\beta_{\mathrm{eq}}\sim 1$, that is, Planck mass relics can account for the entire dark matter. Or, one may regard this temperature as a new upper limit on the reheating temperature \magenta{in this scenario}. 

This mechanism may be similar to Planckian interacting dark matter of Refs. \cite{Garny:2015sjg,Garny:2018grs}. As mentioned there the current observational bound on the tensor-to-scalar ratio \cite{Array:2015xqh,Ade:2015lrj} translates into an upper \blue{bound} on the Hubble parameter of $H\simeq 6.6\times 10^{-6}M_P$. Assuming instantaneous reheating, this corresponds to the reheating temperature of $5.7\times 10^{-4}M_P$ for the effective relativistic degrees of freedom $g_*=106.75$. 
Though our estimations above are admittedly crude, the above mechanism of dark matter creation \blue{is ruled out by} this upper limit. However, \blue{the story would be different if we} generalize the above argument to theories with spatial extra dimensions.

\section{Generalization to theories with extra dimensions}

Let us generalize Eq. (\ref{betaeq}) to cases with extra dimensions as follows. The Planck units can be constructed from $G_D$, $\hbar$ and $c$, where $G_D$ is the generalized gravitational constant in $D$ dimensions (see Ref. \cite{Park:2012fe} for details). The reaction rate is \cyan{$\Gamma=K t_D^{-1}(T/T_D)\exp(-E_D/kT)$} assuming \cyan{$\sigma= K l_D^2$} \magenta{with $K$ being a constant} and \cyan{$v= c$}, and the radiation energy density is $\rho= \blue{(\pi^2g_*/30)}E_Dl_D^{-3}(kT/E_D)^4$. Let us further assume $H^2\blue{
=8\pi G\rho/3c^2}$, where $G$ is the usual, 4-dimensional gravitational constant. Note that this relation may be modified before the big bang nucleosynthesis in a model-dependent way in theories with extra dimensions \cite{ArkaniHamed:1998nn}. Then
\begin{equation}
H\blue{=\left(\frac{4\pi^3g_*}{45}\right)^{1/2}} t_P^{-1}\left(\frac{E_D}{E_P}\right)^2\left(\frac{kT}{E_D}\right)^2
\blue{=\left(\frac{4\pi^3g_*}{45}\right)^{1/2}}t_D^{-1}\left(\frac{E_D}{E_P}\right)\left(\frac{kT}{E_D}\right)^2,
\end{equation}
where $t_DM_D=\hbar c^{-2}=t_PM_P$ was used. Hence we find
\begin{equation}
\beta_{\mathrm{eq}}=\blue{\frac{30K}{\pi^2g_*}\left(\frac{45}{4\pi^3g_*}\right)^{1/2}}\frac{kT/E_D}{kT_{eq}/E_P}\left(\frac{kT}{E_D}\right)^{-4}\exp\left(-\frac{2E_D}{kT}\right).
\end{equation}
Note that replacing $kT/E_D\rightarrow kT/E_P$, we recover Eq. (\ref{betaeq}). Hence, also in this case, $\beta_{\mathrm{eq}}\sim 1$ is realized if the reheating temperature is $kT\sim 0.01 E_D$. In order for relics to serve as dark matter, they have to be confined to the brane, which is the case \cite{Park:2012fe} for black holes in the scenario of Ref. \cite{Randall:1999ee}. Again, this temperature can also be regarded as a new upper limit on the reheating temperature, under the assumption that relics are left over and they stay on the brane.

\blue{So far we} have assumed \blue{the} instantaneous reheating \blue{after} inflation, but our discussions can be generalized to \blue{the case the Universe is dominated by the inflaton field oscillation}
before the radiation-dominated epoch. The energy density of relics created by a moment $t$ during the oscillation phase is
\begin{equation}
a^3(t)\rho(t)=\int n E_D\Gamma(t')a^3(t')dt'=\frac{KE_D}{l_D^3}\int\left(\frac{T}{T_D}\right)^2\exp\left(-\frac{2E_D}{kT}\right)a^3(t')\frac{dt'}{t_D}.
\end{equation}
The energy density of radiation created by the decay of the inflaton with the decay rate $\Gamma_\phi$ is \magenta{\cite{Yokoyama:2004pf}} $\rho_r\sim \red{(2/5)}\Gamma_\phi H^{-1}(H/H_i)^2\rho_i$, where $H_i$ is the Hubble parameter during inflation and $\rho_i$ is the energy density of the inflaton at the beginning of the oscillation phase, which we assume to be $\rho_i\sim \blue{(3/8\pi)}M_P^2H_i^2=\blue{(3/8\pi)}(t_PH_i)^2E_Pl_P^{-3}.$ Assuming instantaneous thermalization, the above $\rho_r$ can be equated with $\rho_r\sim \blue{(\pi^2g_*/30)}E_Dl_D^{-3}(T/T_D)^4$ to obtain the relation $t\sim \blue{(3/\pi^3g_*)}\Gamma_\phi t_P^2\rho_P/\rho_D(T/T_D)^{-4}$, where $\rho_{P,D}=E_{P,D}l_{P,D}^{-3}.$ Then the above integration over time $t$ can be rewritten as an integration over temperature $T$, and the energy density of relics at the reheating is 
\begin{equation}
\rho(t_R)=\blue{\frac{12K}{\pi^3g_*}}\rho_D\frac{t_P^2}{\Gamma_\phi^{-1}t_D}\frac{\rho_P}{\rho_D}\left(\frac{T_R}{T_D}\right)^8\int_{T_R}^{T_{\mathrm{max}}}\exp\left(-\frac{2E_D}{kT}\right)\frac{T_D^{10}dT}{T^{11}}.
\end{equation}
Introducing $x\equiv E_D/kT$ and assuming $1\ll x_{\mathrm{max}}\ll x_R$, the above integration can be approximated as $\int_{x_{\mathrm{max}}}^{x_R}\exp(-2x)x^9dx\sim \blue{2^{-1}}x_{\mathrm{max}}^9\exp(-2x_{\mathrm{max}})$. Noting that $\beta$ starts to decay as $a^{-1}$ at the reheating, $\beta_{\mathrm{eq}}$ in this case turns out to be
\begin{equation}
\beta_{\mathrm{eq}}=\blue{\frac{180K}{\pi^5g_*^2}}\frac{T_R}{T_{\mathrm{eq}}}\frac{t_P^2}{\Gamma_\phi^{-1}t_D}\frac{\rho_P}{\rho_D}\left(\frac{T_R}{T_D}\right)^4\left(\frac{E_D}{kT_{\mathrm{max}}}\right)^9\exp\left(-\frac{2E_D}{kT_{\mathrm{max}}}\right).
\end{equation}
Note that in this case, the abundance of relics is mostly determined by how close $T_{\mathrm{max}}$ is to $T_D$, \magenta{with a much weaker dependence on} $T_R$. The reheating may be defined as the moment when $H^2\blue{=(8\pi/3)} G\rho_r\blue{=A} t_D^{-2}(E_D/E_P)^2(T_R/T_D)^4=\Gamma_\phi^2$, \magenta{where} $\blue{A=4\pi^3g_*/45},$ and this leads to $T_R=\blue{A^{-1/4}}(\gamma H_it_D)^{1/2}(E_P/E_D)^{1/2}T_D$, where $\gamma$ was defined by writing $\Gamma_\phi=\gamma H_i$. On the other hand, $T_{\mathrm{max}}= (\blue{B}\gamma H_i^2t_P^2\rho_P/\rho_D)^{1/4}T_D$, $\blue{B=9/2\pi^3g_*}$. Using these relations, one may find
\cyan{
\begin{align}
\!\!\!\!&\!\!\!\!\!\beta_{\mathrm{eq}}=\frac{180K}{\pi^5g_*^2} A^{-1/4}\left(\gamma H_it_D\frac{E_P}{E_D}\right)^{1/2}\frac{T_P}{T_{\mathrm{eq}}}\gamma H_i\frac{t_P^2}{t_D}\frac{\rho_P}{\rho_D}\left(\gamma H_it_D\frac{E_P}{E_D}\right)^2A^{-1}\left(B\gamma H_i^2t_P^2\frac{\rho_P}{\rho_D}\right)^{-9/4}\exp\left[-2\left(B\gamma H_i^2t_P^2\frac{\rho_P}{\rho_D}\right)^{-1/4}\right]\nonumber\\
&=\frac{180K}{\pi^5g_*^2}\left(\frac{4}{10}\right)^{-5/4}\frac{2\pi^3g_*}{9}\gamma^{5/4}(t_DH_i)^{-1}\frac{T_D}{T_{\mathrm{eq}}}\left(\frac{E_P}{E_D}\right)^{5/2}\left(\frac{t_P}{t_D}\right)^{-5/2}\left(\frac{\rho_P}{\rho_D}\right)^{-5/4}\exp\left[-2\left(B\gamma H_i^2t_D^2\frac{t_P^2}{t_D^2}\frac{\rho_P}{\rho_D}\right)^{-1/4}\right],
\end{align}
}
\magenta{so that}
\begin{equation}
\beta_{\mathrm{eq}}= \blue{C}\gamma^{5/4}(t_DH_i)^{-1}\frac{T_D}{T_{\mathrm{eq}}}\exp
\left\{
-2\blue{(B\gamma)}^{-1/4}\left(
\frac{E_P}{E_D}t_DH_i
\right)^{-1/2}
\right\}, \quad\blue{C=\frac{10\cdot 2^{7/4}5^{9/4}K}{\pi^2g_*}}.
\end{equation}
Then the condition for $\beta_{\mathrm{eq}}=1$ gives
\begin{equation}
\frac{H_i}{M_D}= 4\blue{(B\gamma)}^{-1/2}\frac{E_D}{E_P}\left[\blue{\ln C+}\frac{5}{4}\ln\gamma +\magenta{\ln\left(\frac{H_i}{M_D}\right)^{-1}}+\ln\left(\frac{T_D}{T_{\mathrm{eq}}}\right)\right]^{-2}.\label{condition}
\end{equation}

\cyan{If we require $T_{\mathrm{max}}<T_D$, we find $H_i/M_D<M_D/\sqrt{B\gamma}M_P\simeq 27\gamma^{-1/2}M_D/M_P$, hence this constraint is more stringent than the requirement $H_i/M_D<1$, if one considers $M_D\ll M_P$. From the above equation one gets $H_i/M_D<Ce^2\gamma^{5/4}T_D/T_{\mathrm{eq}}$, which \magenta{would not put any} additional constraint since normally the right hand side here would be larger than unity. Though we may consider $M_D\ll M_P$, if $M_D$ is too small, we find $H_i$ to be too small if we are to create the dark matter by this mechanism, which may be problematic, as discussed in \cite{Kaloper:1998sw}. 
The situation is summarized in Fig. 1, which shows that the above condition for dark matter creation is compatible with both requirements $T_{\mathrm{max}}<M_D$ and $H_i<M_D$. However, one may ensure sufficiently large $H_i$, say, $1$TeV, then we need $10^{12}\mathrm{GeV}<M_D$ from the figure. 
}
\begin{figure}[htbp]
\begin{center}
\includegraphics[width=8cm,angle=0,clip]{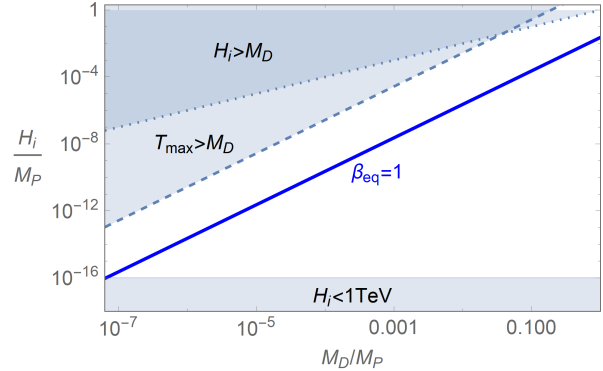}
\caption{The thick line corresponds to Eq. (\ref{condition}), with $\gamma=1$ \magenta{and also $K=1$}. The dashed line is simply $H_i=M_D$. The dotted line is $H_i/M_D=M_D/\sqrt{B\gamma}M_P$ (or $T_{\mathrm{max}}=T_D$, see the texts). \magenta{The lower shaded region corresponds to $H_i<1$TeV, shown for } illustration. }
\label{fig1}
\end{center}
\end{figure}

In models investigated in \blue{Refs. \cite{Giudice:2002vh,Frolov:2002qm}}, the amplitude of tensor perturbations was shown to be written as $k^{3/2}h_k=H/M_{P,\mathrm{eff}}$, where $M_{P,\mathrm{eff}}$ is the effective Planck mass during inflation, which may be different from the current Planck mass. 
 Hence, measurements of B-mode polarizations in the CMB would require that  
 $H$ be sufficiently small relative to the $M_{P,\mathrm{eff}}$, but \red{$M_D$} can in general be smaller than $M_{P,\mathrm{eff}}$. The amplitudes of gravitational waves may change after inflation due to varying $M_{P,\mathrm{eff}}$ \cite{Frolov:2002qm}. \blue{See Refs. \cite{Hiramatsu:2003iz,Hiramatsu:2004aa} for evolution of gravitational waves from inflationary brane world. }
 
 In models considered in Ref. \cite{Im:2017eju}, the upper limit on the reheating temperature was obtained from the condition to avoid strongly coupled gravity. The upper limit thus obtained can be comparable to the 5D gravity scale in examples mentioned there. See also Ref. \cite{ArkaniHamed:1998nn} for several arguments placing the upper limits on the reheating temperature in theories with extra dimensions, but in their models created micro black holes are expected to leave the brane \cite{Park:2012fe}.

\section{Discussion}
There are also other mechanisms of micro black hole formation at high temperatures, which may give contributions similar to the estimations above. One of these is quantum gravitational tunneling of Ref. \cite{Gross:1982cv} (see also Refs. \cite{Kapusta:1984yh} and \cite{Scardigli:2010gm} for a heuristic derivation and Ref. \cite{Hayward:1989jq} for its cosmological implications). The formula for nucleation rate needs to be somewhat modified when relics are left over \cite{Barrow:1992hq}.
Black holes can also arise from thermal fluctuations \cite{PiranWald}, and the efficiency was noted to be less than the above mechanism of quantum gravitational tunneling. 
There is also an analogous process of black hole creation during a de-Sitter phase, with the rate $\sim e^{-T_P^2/12\pi T_{\mathrm{dS}}^2}$ \cite{Ginsparg:1982rs}, with $T_{\mathrm{dS}}=H/2\pi$. Black holes created near the end of inflation would not have experienced substantial dilution. 
It would be interesting to consider micro black hole formation in other modified theories of gravity, as in Refs. \cite{Paul:2005wb,Dialektopoulos:2017pgo}. \blue{Recently it was pointed out that black hole remnants may exist if Starobinsky inflation occurred \cite{anderson}. }


Micro black holes generated by high-energy collisions might lose part of its mass by Hawking radiation before fully reaching the mass of stable relics, and this Hawking radiation could be related to baryogenesis \cite{Alexander:2007gj}.

One of the interesting implications of Planck mass relics as cold dark matter formed in the early Universe would be a relatively small minimum mass of dark matter halos, which we estimate as follows, \blue{based on} Ref. \cite{Schneider:2013ria}. The comoving free streaming scale at the matter-radiation equality $t_{eq}$ is
\begin{equation}
\lambda_{\mathrm{fs}}=\int_{t_R}^{t_{\mathrm{eq}}}\frac{v(t)dt}{a(t)}\sim \frac{2ct_{\mathrm{eq}}T_{\mathrm{eq}}}{a_{\mathrm{eq}}T_R}\ln\frac{T_R}{T_{\mathrm{eq}}},
\end{equation}
assuming the initial velocity of relics at $t_R$ is of the order of the speed of light, and the velocity subsequently decays as $a^{-1}$. Then the minimum mass of halos is $M_{\mathrm{min}}\sim \lambda_{\mathrm{fs}}^3\Omega_m\rho_{\mathrm{cr}}$. For $T_R=1$TeV, $M_{\mathrm{min}}\sim 10^{-15}M_\odot$, and it is even smaller for larger reheating temperature. Hence, the mass function of dark matter halos is expected to continue down to very small masses, which may be observationally tested by methods such as precise pulsar timing measurements in future \cite{Kashiyama:2018gsh}.

\begin{acknowledgments}
TN was partially supported by
JSPS Postdoctoral Fellowships for Research Abroad. The work of JY was supported by JSPS
KAKENHI, Grant JP15H02082 and Grant on Innovative Areas JP15H05888.
\end{acknowledgments}

\end{document}